\documentclass[aps,prd,a4paper,preprintnumbers,amsmath,amssymb,superscriptaddress,twocolumn,floatfix,showpacs]{revtex4}
\usepackage{amssymb}
\usepackage{float,graphicx}
\newcommand{\Oo}{{\cal O}}
\newcommand{\dd}{{\rm d}}

\newcommand{\GeV}{\,{\rm GeV}}

\newcommand{\kpc}{\,{\rm kpc}}
\newcommand{\pc}{\,{\rm pc}}

\newcommand{\eV}{\,{\rm eV}}

\newcommand{\Si}{{\rm Si}}

\newcommand{\etal}{\emph{et al.}\,}
\newcommand{\Pphi}{\mathcal{P}_{\gamma \leftrightarrow \phi}}
\newcommand{\Pbar}{\bar{\mathcal{P}}_{\gamma \leftrightarrow \phi}}
\newcommand{\be}{\begin{eqnarray}}
\newcommand{\ee}{\end{eqnarray}}
\newcommand{\ba}{\left( \begin{array}{ccc}}
\newcommand{\ea} {\end{array} \right)}
\newcommand{\bv}{\left( \begin{array}{c}}
\newcommand{\ev} {\end{array} \right)}
\newcommand{\hatb}[1]{\hat{\mathbf{#1}}}
\begin{document}
\title{The Chameleonic Contribution to the SZ Radial Profile of the Coma Cluster}
\author{Anne-Christine Davis}
\email{a.c.davis@damtp.cam.ac.uk}
\affiliation{Department of Applied Mathematics and Theoretical Physics,
Centre for Mathematical Sciences,  Cambridge CB3 0WA, United Kingdom}
\author{Camilla A.O. Schelpe}
\email{C.A.O.Schelpe@damtp.cam.ac.uk}
\affiliation{Department of Applied Mathematics and Theoretical Physics,
Centre for Mathematical Sciences,  Cambridge CB3 0WA, United Kingdom}
\author{Douglas J. Shaw}
\email{D.Shaw@qmul.ac.uk}
\affiliation{Queen Mary University of London, Astronomy Unit, Mile End Road, London E1 4NS, United Kingdom}
\begin{abstract}
We constrain the chameleonic Sunyaev--Zel'dovich (CSZ) effect in the Coma cluster from measurements of the Coma radial profile presented in the WMAP 7-year results. The CSZ effect arises from the interaction of a scalar (or pseudoscalar) particle with the cosmic microwave background in the magnetic field of galaxy clusters. We combine this radial profile data with SZ measurements towards the centre of the Coma cluster in different frequency bands, to find $\Delta T_{\rm SZ,RJ}(0)=-400\pm40\,\mu\mathrm{K}$ and $\Delta T_{\rm CSZ}^{204\mathrm{GHz}}(0)=-20\pm15\,\mu\mathrm{K}$ (68\% CL) for the thermal SZ and CSZ effects in the cluster respectively. The central value leads to an estimate of the photon to scalar (or pseudoscalar) coupling strength of $g_{\rm eff} \approx (5.2 \text{--} 23.8) \times 10^{-10}\,{\rm GeV}^{-1}$, while the 95\% confidence bound is estimated to be $g_{\rm eff} \lesssim (8.7 \text{--} 39.4) \times 10^{-10}\,{\rm GeV}^{-1}$.
\end{abstract}
\maketitle

\section{Introduction}
There has been a great deal of interest over the past 25 years or so in the mixing of axion-like particles (ALPs) with photons in the presence of magnetic fields; for example Refs. \cite{Das05, Raffelt88, Harai92, Burrage08} or see Ref. \cite{ALPrev} for a recent review. ALPs are a generalisation of the Peccei-Quinn axion originally introduced to solve the strong CP problem \cite{Peccei77}. They refer to any light scalar or pseudoscalar with a linear coupling to the electromagnetic (EM) tensor, and have two free parameters: the axion mass and the EM coupling strength. For more details of the axion and the experimental bounds placed on its parameters see for example \cite{Masso03,ALPrev}. More exotic ALPs have been postulated, such as the chameleon scalar field developed by Khoury and Weltman \cite{Khoury04}. The chameleon scalar field has a density-dependent mass, making it an attractive candidate for dark energy. In high-density environments such as on Earth the chameleon is heavy and evades fifth-force laboratory searches despite a strong coupling to matter; in low-density environments the chameleon is light and can drive the cosmic acceleration. Bounds placed on the axion in the laboratory are evaded by the chameleon and so the chameleon-photon coupling strength is less well constrained. The strongest bounds come from mixing in large-scale astrophysical magnetic fields, see for example Refs. \cite{BurrageSN,Burrage08,Davis09}. It was pointed out in \cite{Burrage08} that in sparse astrophysical environments, such as that found in galaxies or galaxy clusters, the mixing between the chameleon field and photons is indistinguishable from the mixing of any very light ALP. In our analysis we assume full generality as to the nature of the ALP but the resulting constraints will be most applicable to the chameleon scalar field. A detailed discussion of the allowed chameleon parameters is given in Ref. \cite{MotaShaw}. The best direct upper-bounds on the matter coupling $1/M$ come simply from the requirement that the chameleon field makes a negligible contribution to standard quantum amplitudes, roughly $1/M \lesssim 10^{-4} \,{\rm GeV}^{-1}$ \cite{MotaShaw, BraxLight}. Stronger constraints have been derived on the photon coupling $g_{\rm eff} = 1/M_{\rm eff}$. Laser-based laboratory searches for general ALPs, such as PVLAS \cite{PVLAS}, and chameleons, such as GammeV \cite{GammeV}, give $g_{\rm eff} \lesssim 10^{-6}\,{\rm GeV}^{-1}$.  Constraints on the production of starlight polarization from the mixing of photons with chameleon-like particles in the galactic magnetic field provide an upper-bound of $g_{\rm eff} < 9 \times 10^{-10}\,{\rm GeV}^{-1}$ \cite{Burrage08}.  

In \cite{Davis09} we predicted the existence of a chameleonic Sunyaev-Zel'dovich (CSZ) effect arising from the mixing between a light scalar (or pseudoscalar) field and the cosmic microwave background (CMB) radiation in the magnetic field of galaxy clusters, with particular reference to the chameleon model. The scalar-photon interaction causes a fraction of the photons to be converted along the path, decreasing the overall photon intensity. The effect is similar to the thermal Sunyaev-Zel'dovich (SZ) effect arising from scattering of photons off electrons in the cluster atmosphere, but with a significantly different frequency dependence. In \cite{Davis09} we compared predictions of the thermal SZ and CSZ effects in the Coma cluster of galaxies to the measured SZ decrement towards the centre of the cluster in a number of frequency bands. The results constrained the scalar-photon coupling strength to be, $g_{\rm eff} < (0.72-22) \times 10^{-9}\,{\rm GeV}^{-1}$, depending on the model that is assumed for the cluster magnetic field structure. 

In \cite{Davis09} we proposed that the constraints would be stronger if the radial profile of the SZ effect could be analysed. The thermal SZ effect scales with the integrated electron density in the cluster and decreases rapidly as the electron density decreases towards the edge of the cluster. The chameleonic SZ effect, by contrast, depends on the ratio of the magnetic field to the electron density, which introduces an extended wing to the predicted radial profile. This additional power in the SZ signal at larger radii is a distinctive signature. Until recently the only analysis has been of the stacked SZ profiles of galaxy clusters at known positions on the sky, suggesting a discrepancy between the SZ intensity decrement as measured by WMAP and the expected thermal SZ signal based on the X-ray emission data \cite{ROSATvsWMAP,Komatsu10}. This discrepancy takes the form of a greater than expected SZ signal at large radii. One possibility is that a simple beta-model is inadequate for describing the radial profile of the electron density distribution in the cluster and that a more sophisticated model is required, but another more exciting possibility is the presence of a chameleon scalar field. 

The recent WMAP 7-year SZ measurements of the Coma cluster radial profile \cite{Komatsu10} are the first detailed observations of the SZ profile around a cluster to date. We calculate the CSZ contribution to the signal in the combined $V+W$ band extracted from the WMAP measurements; we believe it is beyond the scope of this paper to perform a full analysis of the raw WMAP data contaminated by noise and CMB anisotropies. 

This paper is organised as follows: in section \ref{sec:ALPmixing} we present the calculations for the mixing between ALPs and photons in the magnetic field of galaxy clusters. In section \ref{sec:ComaCluster} we discuss measurements of the magnetic field structure and SZ effect in the Coma cluster. The results of our analysis are in section \ref{sec:CSZeffect} and we conclude with a discussion of the results in section \ref{sec:Conclusions}. Appendix \ref{appendix:A1} contains the derivation for the probability of conversion between photons and ALPs in galaxy clusters. Appendix \ref{appendix:A2} outlines our simplistic approach to calculating the contribution of the CSZ effect to the $V+W$ band in the WMAP data.

\section{Mixing between Photons and Axion-like Particles in Cluster Magnetic Fields \label{sec:ALPmixing}}
The interaction Lagrangians describing coupling between photons and scalar- or pseudoscalar-ALPs are,
\begin{eqnarray}
\mathcal{L}_{\rm scalar} & = & -\frac{1}{4}g_{\rm eff}\phi F_{\mu\nu}F^{\mu\nu}\nonumber\\
\mathcal{L}_{\rm pseudoscalar} & = & +\frac{1}{4}g_{\rm eff}\phi F_{\mu\nu}\tilde{F}^{\mu\nu},\nonumber
\end{eqnarray}
where $\tilde{F}_{\mu\nu}=\epsilon_{\mu\nu\alpha\beta}F^{\alpha\beta}/2$ is the dual of the field strength tensor. The difference between the scalar and pseudoscalar interaction determines which photon polarization state is involved in mixing. The modification to the intensity of any radiation is therefore identical for either particle but the modification to the polarization state is particle specific.

The resulting field equations for this interaction are,
\begin{eqnarray}
\Box \phi & =  m_{\phi}^2 \phi & +\frac{1}{4}g_{\rm eff}F_{\mu\nu}F^{\mu\nu},\label{field1}\\
&&\left(-\frac{1}{4}g_{\rm eff}F_{\mu\nu}\tilde{F}^{\mu\nu}\right),\nonumber
\end{eqnarray}
and, 
\begin{eqnarray}
\nabla_{\nu}F^{\mu\nu} &= J^{\mu}&-g_{\rm eff}\nabla_{\nu}\left(\phi F^{\mu\nu}\right),\label{field2}\\
&&\left(+ g_{\rm eff}\nabla_{\nu}(\phi \tilde{F}^{\mu\nu})\right),\nonumber
\end{eqnarray}
where the bracketed term is for a pseudoscalar as opposed to a scalar ALP. $J^{\mu}$ is the background electromagnetic 4-current, such that $\nabla_{\mu}J^{\mu}=0$, and $m_{\phi}$ is the mass of the ALP. The Lagrangian for the chameleon scalar field must include the coupling between the scalar field and matter. However in sparse astrophysical environments the varying properties of the chameleon can be encapsulated in its mass $m_{\phi}$ to give the same field equations as Eqns. (\ref{field1}) and (\ref{field2}). See \cite{Davis09} for a full derivation. 

We follow a similar procedure to that given in Refs. \cite{Raffelt88,Davis09,Burrage08,Das05} to determine the evolution of the ALP and photon fields, when propagating through a background magnetic field $\mathbf{B}(\mathbf{x})$. We expand the fields as perturbations about their background values, $\varphi = \phi - \bar{\phi}$ and $a^{\mu}=A^{\mu}-\bar{A}^{\mu}$, where $F_{\mu\nu}=\partial_{\mu}A_{\nu}-\partial_{\nu}A_{\mu}$. Ignoring terms that are second order in the perturbations, we find
\begin{eqnarray}
-\partial_t^2 \phi+\nabla^2 \phi &\simeq m_{\phi}^2 \phi & + g_{\rm eff} \mathbf{B}.(\boldsymbol{\nabla}\times\mathbf{a}), \label{eq:mixing1}\\
&&\left(+ g_{\rm eff} \mathbf{B}.(\partial_t\mathbf{a})\right), \nonumber
\end{eqnarray}
and, 
\begin{eqnarray}
-\partial_t^2 \mathbf{a}+\nabla^2 \mathbf{a} &\simeq \omega_{\rm pl}^2 \mathbf{a} &+ g_{\rm eff} \left(\boldsymbol{\nabla}\phi\times\mathbf{B}\right),\label{eq:mixing2} \\
&&\left(- g_{\rm eff} \mathbf{B} (\partial_t \phi)\right),\nonumber
\end{eqnarray}
where again the bracketed term applies for pseudoscalars. The plasma frequency, $\omega_{\rm pl}^2=4\pi \alpha_{\rm EM} n_{\rm e}/m_{\rm e}$, arises from interactions between the photons and electrons of number density $n_{\rm e}$ in the plasma, and acts as an effective photon-mass. 

The probability of conversion between photons and chameleons after passing through the magnetic field of a galaxy cluster was derived in \cite{Davis09}. The details are reproduced here in Appendix \ref{appendix:A1}, and have been extended to apply generally to scalar- and pseudoscalar-ALPs. The derivation assumes weak-mixing between the photons and ALPs (valid for CMB photons propagating through galaxy clusters), and requires the ALP mass to be below the plasma frequency in the cluster. For the case of the Coma cluster this imposes $m_{\phi} \lesssim 10^{-12}\mathrm{eV}$. It was shown in \cite{Davis09} that for CMB photons travelling through the intracluster (IC) medium the mass of the chameleon scalar field satisfies this requirement.

The magnetic fields of galaxy clusters have been measured to have a regular component $\mathbf{B}_{\rm reg}(\mathbf{x})$, with a typical length scale of variation similar to 
the size of the object, and a turbulent component $\delta\mathbf{B}(\mathbf{x})$ which undergoes $\Oo(1)$ variations and reversals on much smaller scales. Instead of considering a simple cell model for the magnetic field fluctuations, common in both the 
literature concerning photon-ALP conversion in galaxy and cluster magnetic fields and in that concerning the measurement of those fields, we consider a more realistic model for $\mathbf{B}$ assuming a spectrum of fluctuations running from some very large scale (e.g. the scale of the galaxy or cluster) down to some very small scale ($\ll L_{\rm coh}$).  We describe these fluctuations by a correlation function $R_{{\rm B}\, ij}(\mathbf{x}) = \left\langle \delta B_{i}(\mathbf{y})\delta B_{j}(\mathbf{x}+\mathbf{y})\right\rangle$; here the angled brackets indicate the expectation of the quantity inside them.  This description of the magnetic field is compatible with the requirement that $\nabla \cdot \mathbf{B}=0$ \cite{Ensslin03}. Assuming that the fluctuation spectrum is isotropic we have:
$$
R_{{\rm B}\, ij}(\mathbf{x}) = \frac{1}{3}\delta_{ij} R_{{\rm B}}^{\perp}(x) + \hat{x}_{i}\hat{x}_{j}R_{\rm B}^{\parallel}(x),
$$
where $R_{\rm B}^{\parallel}(x)$ is related to $R_{{\rm B}}^{\perp}(x)$ by imposing $\nabla\cdot\mathbf{B}=0$.  The photon-scalar conversion probability only depends on the components of the magnetic field that are perpendicular to direction of the path, and so do not depend on $R_{\rm B}^{\parallel}(x)$.  We define $P_{{\rm B}}(k)$ to be the power spectrum associated with $R_{{\rm B}}^{\perp}(x)$.   We allow for a similar spectrum of fluctuations in the electron number density, $n_{\rm e}(z)=\bar{n}_{\rm e}+\delta n_{\rm e}(z)$, with auto-correllation function $R_{N}(x)$ and associated power spectrum $P_{\rm N}(k)$.

The power spectra, $P_{{\rm B}}(k)$ and $P_{\rm N}(k)$, have not been measured for the Coma cluster, but we draw on observations of our own galaxy and other clusters to infer their probable form \cite{Han04,Ensslin03,Murgia04,Davis09}.  These measurements are generally consistent with a Kolmogorov-type turbulence model, but only probe the power spectrum at spatial scales larger than a few kiloparsecs.  The dominant contribution to the probability of photon-ALP conversion in galaxy clusters is 
sensitive to $P_{{\rm B}}(k)$ and $P_{\rm N}(k)$ on scales $k^{-1} \lesssim 10^{-3}\textrm{--}0.1\,{\rm pc}$.  There are no observations of the form or magnitude of $P_{\rm B}(k)$ and $P_{\rm N}(k)$ on such small scales. According to Kolmogorov's 1941 theory of turbulence, the power spectrum of three dimensional turbulence on the smallest scales is universal and has $P(k) \propto k^{-11/3}$. We therefore estimate $P_{\rm B}(k)$ and $P_{\rm N}(k)$ in the Coma cluster on small scales  by assuming we have three dimensional Kolmogorov turbulence.

We find that the dominant contribution to the average probability of photon-ALP conversion in a galaxy cluster is (Eq. (\ref{eq:Pbar}), Appendix \ref{appendix:A1}),
\be
\Pbar &\approx& \frac{3}{40}(0.27\text{--}0.45) g_{\rm eff}^{2}\left(\frac{\alpha_{\rm EM}\bar{n}_{\rm e}}{2\pi m_{\rm e}}\right)^{-5/3} \nu^{5/3}L_{\rm B}^{-2/3} L \nonumber \\
&& \cdot\left(B_{\rm reg}^{2}\frac{I_{\rm N}^2\left(I_{\rm N}-1\right)}{2}+ \left\langle \delta \mathbf{B}^2\right\rangle \frac{I_{\rm N}^2 \left(2I_{\rm N}-1\right)}{3}\right), \nonumber \\
&&\label{Pbar}
\ee
where $ I_{\rm N} \equiv 1+\left\langle \delta n_{\rm e}^2/\bar{n}_{\rm e}^2\right\rangle$, $\nu$ is the frequency of the radiation, $L$ is the path length through the magnetic region, and $L_{\rm B}$ is the coherence length of the magnetic field which is defined later in Eq. (\ref{lengths}). The range depends on the normalisation for the power spectra $P_{\rm B}(k)$ and $P_{\rm N}(k)$.
The resulting modification to the CMB intensity is,
\be
\frac {\Delta I}{I_0} = -\Pbar, \nonumber
\ee
which holds equally for both scalar- and pseudoscalar-ALPs.

\section{The Coma Cluster\label{sec:ComaCluster}}
The nearby Coma cluster of galaxies is unique in having detailed measurements of both its magnetic field structure and the SZ effect. In this section we discuss these measurements.

\subsection{Magnetic field}
The magnetic field at the centre of the Coma cluster was determined by Feretti \textit{et al.} \cite{Feretti95} from observations of Faraday rotation measures (RMs) of an embedded source. The RM corresponds to a rotation of the polarization angle of an intrinsically polarised radio source. The RM along a line of sight in the $\hatb{z}$ direction of a source at $z=z_{\rm s}$ is given by:
\be
{\rm RM}(z_{\rm s}\hatb{z}) = a_{0}\int_{0}^{z_{\rm s}} n_{\rm e}(x \hatb{z})B_{\parallel}(x \hatb{z}) \dd x, \nonumber
\ee
where $a_{0} = \alpha_{\rm EM}^3/\pi^{1/2} m_{\rm e}^2\,$, $B_{\parallel}$
is the magnetic field along the line of sight, and the observer is at
$\mathbf{x}=0$. For a uniform magnetic field over a length $L$ along the line of sight and in a region of constant density, we expect $\langle {\rm RM}\rangle=a_{0} B_{\parallel}n_{\rm e}L$. However, the dispersion of Faraday RMs from extended
radio sources suggests that this is not a realistic model for IC
magnetic fields. Both the dispersion of the measured RMs and simulations suggest that IC magnetic fields typically have a sizable component that is tangled on scales smaller than the cluster. The simplest model for a tangled magnetic field
is the cell model where the
cluster is divided up into uniform cells with sides of length $L_{\rm
  coh}$.  The total magnetic field strength is assumed to be the same
in every cell, but the orientation of the magnetic field in each cell
is random. In the cell model the RM is given by a
Gaussian distribution with zero mean, and variance
\be
\left \langle {\rm RM}^2 \right \rangle \sim \frac{a_{0}^2 L_{\rm coh}} {2} \int_{0}^{L} \left \langle \mathbf{B}^2(z)\right\rangle n_{\rm e}^2(z) \dd z,\nonumber
\ee
where $\left \langle \mathbf{B}^2\right\rangle$ is the average value of $\mathbf{B}^2$ along the line of sight.
A more realistic model for the magnetic field is to assume a power spectrum of fluctuations.  In this model one finds \cite{Murgia04},
\be
\left \langle {\rm RM}^2 \right \rangle \sim \frac{a_{0}^2  L_{\rm B} }{2} \int_{0}^{L} \left \langle \mathbf{B}^2(z) \right\rangle n_{e}^2(z) \dd z\,, \nonumber
\ee
where following Ref. \cite{Murgia04} the correlation length scales, $L_{\rm B}$ and $L_{\rm N}$, for the magnetic and electron density fluctuations respectively are defined,
\be
L_{\rm B/N} = \frac{ \int_{0}^{\infty} k\dd k \,P_{\rm B/N}(k)}{2 \int_{0}^{\infty} k^2\dd k \, P_{\rm B/N}(k)}. \label{lengths}
\ee
To match the same observations with the same total field strength, $L_{\rm B}$ in the power spectrum model should be equal to $L_{\rm coh}$ in the cell model. 

Assuming a cell model for the Coma cluster magnetic field in combination with a uniform magnetic field, Feretti \textit{et. al.} \cite{Feretti95} detected a uniform component of strength $0.2\pm0.1\,\mu\mathrm{G}$ coherent on scales of $200\,\mathrm{kpc}$ and a tangled component of strength $8.5\pm1.5\,\mu\mathrm{G}$ coherent on much shorter scales of $1\,\mathrm{kpc}$ extending across the core region of the cluster.

\subsection{Radial Profile\label{sec:radialprofile}}

The electron density distribution plays a crucial role in determining both the radial profile of the thermal SZ effect and that of the CSZ effect. 
The electron density of cluster atmospheres can be determined from
X-ray surface brightness observations, for example using ROSAT, which
are generally fitted to a simple beta profile (see Ref. \cite{Govoni04}): 
\be
n_{e}(r)=n_{0}\left(1+\frac{r^{2}}{r_{\rm c}^{2}}\right)^{-3\beta/2}, \label{beta_model}
\ee
where $n_0$ is the electron density at the cluster centre, $r_{\rm c}$ is the core radius (a free parameter fitted to the X-ray observations) and $\beta\sim0.6\text{--}1$ in general. For the Coma cluster the best fitting parameters are \cite{Briel92},
\be
\beta &=& 0.75 \pm 0.03, \nonumber\\
\theta_{\rm c}=r_{\rm c}/D_{\rm A} &=& 10.5 \pm 0.6\;\mathrm{arcmin}, \nonumber
\ee
where $D_{\rm A}=67\,h^{-1}\,\mathrm{Mpc}$ is the angular diameter distance to the Coma cluster.
The average electron density at the centre of the cluster has been measured at $4\times10^{-3}\mathrm{cm}^{-3}$.

In a similar fashion the magnetic field
is expected to decline with radius and the coherence length to increase. Magneto-hydrodynamic (MHD) simulations of
galaxy cluster formation suggest that $ \left \langle \mathbf{B}^2 \right\rangle^{1/2} \propto \left \langle n_{e} \right \rangle^{\eta}$ with $\eta=1$
while observations of Abell 119 suggest $\eta=0.9$ \cite{Dolag}.
Two simple theoretical arguments for the value of $\eta$, are (1) that the magnetic field is frozen-in and hence $\eta = 2/3$, or (2) the total magnetic energy, $\propto \left\langle \mathbf{B}^2\right\rangle$, scales as the thermal
energy, $\propto n_{\rm e}$, and hence $\eta = 1/2$ \cite{Govoni04}. Faraday RM observations of clusters \cite{Clarke01} have found that the magnetic field extends as far as the ROSAT-detectable
X-ray emission (coming purely from the electron density distribution). Simulations have also shown that the coherence length increases with distance from the cluster core, however the radial scaling of $L_{\rm coh}$ is uncertain. We make the simple ansatz that, at least inside the virial radius, $L_{\rm coh}\propto n_{\rm e}^{-\gamma/3\beta}$ for some $\gamma$, so that 
\be
L_{\rm coh}=L_{\rm coh0}\left(1+\frac{r^2}{r_{\rm c}^2}\right)^{\gamma/2},\nonumber
\ee
where $L_{\rm coh0}$ is the coherence length in the centre of the cluster. Based solely on dimensional grounds, estimated values for $\gamma$ are: $\gamma=\beta$ so that $L_{\rm coh}\propto n_{\rm e}^{-1/3}$, or $\gamma=1$ so that $L_{\rm coh}\propto r$ for large $r$. The simulations of Ref. \cite{Ohno03}, suggest that for large $r$, $L_{\rm coh}$ roughly scales as $r$ inside the virial radius.

\subsection{Sunyaev-Zel'dovich measurements\label{SZmeasurements}}

The thermal SZ effect arises from inverse-Compton scattering of CMB photons with free
electrons in a hot plasma. The scattering process redistributes the
energy of the photons causing a distortion in the frequency spectrum
of the intensity. This is observed as a change in the temperature of the CMB radiation in the direction of galaxy clusters \cite{Birkinshaw99,Komatsu10},
\be
\frac{\Delta T_{\mathrm{SZ}}}{T_{0}}= g_{\rm SZ}(\nu)\frac{\sigma_{\rm T}}{m_{\rm e}}\int P_{\rm e}(l)\mathrm{d}l\,,\label{eq:SZ effect}
\ee
where the integral is along the line of sight. The frequency dependence of the effect is contained in $g_{\rm SZ}(\nu)\equiv x\coth\left(\frac{x}{2}\right)-4$
where $x=h\nu/k_{B}T_{0}$. At low frequencies the SZ effect is seen as a decrement in the expected
temperature from the CMB, while at high frequencies it is seen as a
boost in the temperature. The electron pressure, $P_{\rm e}=n_{\rm e}k_{\rm B}T_{\rm e}$, varies with the distribution in the cluster, where $T_{\rm e}$ is the temperature of the electrons in the plasma. The SZ effect is also commonly expressed in terms of the optical depth, defined by $\tau_{0}=\int\sigma_{\rm T}n_{\rm e}(l)\mathrm{d}l$.

Measurements of the CMB temperature decrement towards the centre of the Coma cluster
were assimilated and analysed in \cite{Batistelli03}. These measurements are summarized in Table \ref{tab:ComaSZ}. Assuming only a thermal SZ contribution the optical depth was found to be $4.7\times10^{-3}$. This assumed an electron temperature of $k_{\rm B}T_{\rm e}=8.2\,\mathrm{keV}$. 
\begin{table}
\begin{centering}
\begin{tabular}{|c|c|c|c|}
\hline 
Experiment & $\nu/\mathrm{GHz}$  & $\delta\nu/\mathrm{GHz}$ & $\Delta T_{\mathrm{SZ}}/\mu K$\tabularnewline
\hline
\hline 
OVRO & 32.0 & 6.5 & $-520\pm83$\tabularnewline
\hline 
WMAP & 60.8 & 13.0 & $-240\pm180$\tabularnewline
\hline 
WMAP & 93.5 & 19 & $-340\pm180$\tabularnewline
\hline 
MITO & 143 & 30 & $-184\pm39$\tabularnewline
\hline 
MITO & 214 & 30 & $-32\pm79$\tabularnewline
\hline 
MITO & 272 & 32 & $172\pm36$\tabularnewline
\hline
\end{tabular}
\end{centering}
\caption{\label{tab:ComaSZ}SZ measurements of the Coma cluster core in different frequency bands.}
\end{table}

An alternative parameterisation of the thermal SZ effect is to consider the central temperature decrement in the Rayleigh--Jeans (RJ) limit ($\nu \rightarrow 0$). Separating out the frequency and radial dependence, and assuming a beta model for the electron distribution (Eq. (\ref{beta_model})), we have 
\be
\Delta T_{\mathrm{SZ}}(\theta)=\Delta T_{\mathrm{SZ,RJ}}(0) \left[1+\frac{\theta^2}{\theta_{\rm c}^2}\right]^{(1-3\beta)/2} \frac{g_{\rm SZ}(\nu)}{-2}, \nonumber
\ee
where the radius $r$ from the centre of the cluster to some
point in the cluster atmosphere has been approximated by $r^{2}\approx(D_{A}\theta)^{2}+l^{2}$. Here $\theta$ is the angle subtended from the cluster
centre to the point and $l$ is the radial distance of the point from
the cluster centre along the line of sight. Within this parameterisation the inferred optical depth for the Coma cluster, $\tau_0\approx 4.7\times10^{-3}$, corresponds to a central temperature decrement in the RJ limit of $-420\,\mu\mathrm{K}$. 

\begin{figure}[htb!]
\begin{centering}
\includegraphics[width=7.5cm]{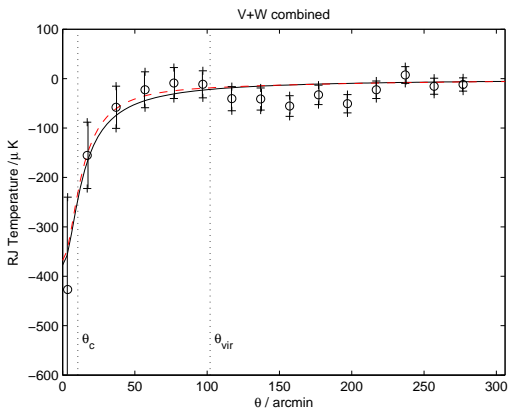}
\end{centering}

\caption{\label{Fig:ComaRP} Measurements of the cleaned SZ signal for the Coma cluster in the combined $V+W$ band, plotted in units of the Rayleigh--Jeans (RJ) temperature, courtesy of Eiichiro Komatsu \cite{Komatsu10}. The solid black line is the best fit thermal SZ model assuming a beta profile for the electron distribution with no CSZ contribution. The dashed blue line is the best fit model assuming both thermal SZ and CSZ contributions.}

\end{figure}

Detailed measurements of the CMB temperature in 43 radial bins centered on the Coma cluster have recently been released by the WMAP collaboration \cite{Komatsu10}. Assuming a beta model for the electron density distribution, they find a central temperature decrement in the Rayleigh--Jeans limit of $-381\pm126\,\mu\mathrm{K}$ from the data extracted in the $V$ band ($60\,\mathrm{GHz}$), and $-523\pm127\,\mu\mathrm{K}$ from the $W$ band ($90\,\mathrm{GHz}$). This discrepancy is most likely caused by contamination from the primary CMB anisotropies. The data is processed to eliminate this CMB signal by combining the $V$ and $W$ bands, to give $\Delta T_{\rm SZ,RJ}(0) = -377\pm105\,\mu\mathrm{K}$ (68\% CL). This suggests a CMB anisotropy signal of the order $-100\,\mu\mathrm{K}$ coinciding with the centre of the Coma cluster. The extracted data in the combined $V+W$ band is reproduced in Fig. \ref{Fig:ComaRP}.

\section{The Chameleonic Sunyaev-Zel'dovich Effect in the Coma Cluster\label{sec:CSZeffect}} 

We saw in section \ref{sec:ALPmixing} that the average conversion rate of photons into ALPs after passing through the magnetic field of a galaxy cluster was given by Eq. (\ref{Pbar}). For the Coma cluster the regular component of the magnetic field has been measured to be much smaller than the tangled component (true of most cluster magnetic fields), and so the conversion probability is approximately, 
\be
\Pbar &\approx& \frac{1}{40}(0.27\text{--}0.45) g_{\rm eff}^{2}\left(\frac{\alpha_{\rm EM}\bar{n}_{\rm e}}{2\pi m_{\rm e}}\right)^{-5/3} \nu^{5/3}L_{\rm B}^{-2/3} L \nonumber \\
&& \cdot\left\langle \delta \mathbf{B}^2\right\rangle I_{\rm N}^2 \left(2I_{\rm N}-1\right). \nonumber
\ee
A reasonable value for $I_{\rm N} \equiv 1+\left\langle \delta n_{\rm e}^2/\bar{n}_{\rm e}^2\right\rangle$ based on electron fluctuations in our own galaxy is $I_{\rm N}\sim 1\text{--}2$. The photon-ALP conversion causes a decrement in the CMB temperature towards galaxy clusters, with
\[
\frac{\Delta T_{\mathrm{CSZ}}}{T_0}=\frac{e^{-x}-1}{x}\bar{P}_{\gamma\leftrightarrow\phi}\,,\]
where $x=h\nu/k_{\rm B}T_{0}$. The above equation assumes the average magnetic field, electron density and coherence length are constant along the path length $L$. However if the fields vary slowly over the path length, then the effect will be proportional to the integral along the line of sight: 
\be
\bar{P}_{\gamma\leftrightarrow\phi}\propto\int\mathrm{d}l\left\langle \delta \mathbf{B}^2\right\rangle \bar{n}_{\mathrm{e}}^{-5/3}L_{\mathrm{B}}^{-2/3}. \nonumber
\ee 
Assuming a beta model for the electron density in the cluster and the radial scaling for the magnetic field and coherence length discussed in section \ref{sec:radialprofile}, we have
\be
\Delta T_{\mathrm{CSZ}}(\theta) &\approx& \Delta T_{\mathrm{CSZ}}^{\rm 204\,GHz}(0) \frac{x_0}{x}\frac{e^{-x}-1}{e^{-x_0}-1}\left(\frac{\nu}{\nu_0}\right)^{5/3} \nonumber \\
&& \cdot\left(1+\frac{\theta^2}{\theta_{\rm c}}\right)^{\frac{p}{2}} \frac{I_{y}(\theta)}{I_{y}(0)}, \label{eq:DeltaCSZ} 
\ee
where $\nu_{0}=204\,\mathrm{GHz}$ and we have defined $p=1-2\gamma/3-6\eta\beta+5\beta$, and 
\be 
I_{y}\equiv \intop_{0}^{\infty}\left(1+y^2\right)^{\frac{p-1}{2}}w(y)\mathrm{d}y\,, \nonumber
\ee
with $y^2 = (l/r_{\rm c})^2\left(1+\frac{\theta^2}{\theta_{\rm c}^2}\right)^{-1}$. We assume the turbulent cluster field dies off very quickly beyond some characteristic radius $r_{{\rm max}}$, due to $L_{{\rm B}}$ growing very quickly or $\left\langle \delta\mathbf{B}^{2}(z)\right\rangle $
decreasing much faster than $\bar{n}_{{\rm e}}^{\eta}$. A reasonable assumption for $r_{{\rm max}}$ is to approximate it by the virial radius $r_{{\rm vir}}$ \cite{Davis09}. Thus we have introduced a window function, $w=\exp(-r^{2}/r_{{\rm vir}}^{2})$, which corresponds to
\[
w(y)=\exp\left(-\frac{\theta^{2}+(\theta_{c}^{2}+\theta^{2})y^{2}}{\theta_{{\rm vir}}^{2}}\right),\]
where $\theta_{{\rm vir}}=r_{{\rm vir}}/D_{A}$. The magnitude of the CSZ effect at the centre of the cluster at $204\,\mathrm{GHz}$ is given by, 
\be
&\frac{\Delta T_{\mathrm{CSZ}}^{\rm 204\,GHz}(0)}{T_0} \approx  -\left(0.13\text{--}2.69 \right)\times 10^{-7}g_{10}^2\left(\frac{\left\langle \delta \mathbf{B}_0^2\right\rangle^{1/2}}{8.5\,\mu{\rm  G}}\right)^2 & \nonumber\\
&\left(\frac{\bar{n}_{\rm e0}}{4 \times 10^{-3}\,{\rm cm}^{-3}}\right)^{-5/3} \left(\frac{r_{\rm c}}{150 \kpc}\right)\left(\frac{L_{\rm B0}}{1 \kpc}\right)^{-2/3},& \label{P204}
\ee
where $g_{10} \equiv g_{\rm eff}/(10^{-10}\GeV^{-1})$.

The WMAP measurements of the Coma cluster radial profile discussed in section \ref{SZmeasurements} constrain the degree of photon-ALP mixing in the Coma cluster. We calculate the contribution of the CSZ effect to the combined $V+W$ band data. This assumes a simplistic approach to the processing technique of Komatsu \textit{et al.} \cite{Komatsu10} in combining the $V$ and $W$ bands to eliminate contamination from the CMB anisotropies. This is detailed in Appendix \ref{appendix:A2}. We predict the total signal to be,
\be
\Delta T_{\rm RJ}^{V+W}(\theta)&=&\Delta T_{\rm SZ,RJ}(\theta) \nonumber \\
&& +\Delta T_{\rm CSZ}^{204\,\mathrm{GHz}}(\theta)\left(\frac{-2(g_{\rm CSZ}^{V}-g_{\rm CSZ}^{W})}{g_{\rm CSZ}(\nu_0)(g_{\rm SZ}^{V}-g_{\rm SZ}^{W})}\right),\nonumber
\ee
where $g_{\rm CSZ}(\nu)\equiv \nu^{5/3}(e^{-x}-1)/x$.  

Comparing our predictions, $T_{\mathrm{pred}}^{(i)}$, to the extracted WMAP data in the combined $V+W$ band, $T_{\mathrm{expt}}^{(i)}$, over the 43 radial bins of the Coma cluster, we maximise
the likelihood $L$ defined by, \[
-2\log L=\sum_{i=1}^{43}\left(\frac{T_{\mathrm{expt}}^{(i)}-T_{\mathrm{pred}}^{(i)}}{\sigma_{\mathrm{expt}}^{(i)}}\right)^2,\]
where $\sigma_{\mathrm{expt}}^{(i)}$ are the errors in the WMAP measurements. We take $\Delta T_{\rm SZ,RJ}(0)$ and $\Delta T_{\rm CSZ}^{204\,\mathrm{GHz}}(0)$ to be the free parameters in the theory and assume no prior knowledge of their distributions. We marginalise over the remaining model parameters $\left(\beta,\,\theta_{\rm c},\,\gamma,\,\eta,\,\theta_{\rm vir}\right)$, and assume the following priors on their distributions:
\be
\beta &\sim& \mathcal{N}\left(0.75,0.03\right), \nonumber \\
\theta_{\rm c} &\sim& \mathcal{N}\left(10.5,0.6\right)\;{\rm arcmin}, \nonumber \\
\gamma &\sim& \mathcal{U}\left(0.6,1\right), \nonumber \\
\eta &\sim& \mathcal{U}\left(0.5,1\right), \nonumber \\
\theta_{\rm vir} &\sim& \mathcal{N}\left(102,20\right)\;{\rm arcmin}, \nonumber 
\ee
where $\mathcal{N}\left(\mu,\sigma\right)$ is a normal distribution with mean $\mu$ and standard deviation $\sigma$; and $\mathcal{U}\left(a,b\right)$ is a uniform distribution over the range $a$ to $b$. These priors come from the experimental measurements and theoretically motivated values discussed in section \ref{sec:radialprofile}. The virial radius of the Coma cluster is $1.99\pm0.22\,h^{-1}\mathrm{Mpc}$ \cite{Kubo08} which, coupled with $D_{\rm A}\approx 67\,h^{-1}\mathrm{Mpc}$, gives the above prior on $\theta_{\rm vir}$. To find the confidence limits on our results we assume that $
\chi^{2}=-2\log\left(L/\hat{L}\right)$ follows a $\chi_1^{2}$ distribution, where $\hat{L}$
is the maximised likelihood over the parameter space. 

\begin{figure}[htb!]
\begin{centering}
\includegraphics[width=7.5cm]{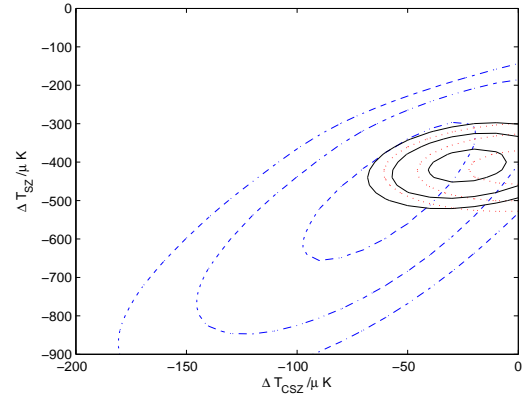}
\end{centering}
\caption{\label{Fig:CLs}Confidence limits on the parameter space of $\Delta T_{\rm SZ,RJ}(0)$ and $\Delta T_{\rm CSZ}^{204\,\mathrm{GHz}}(0)$ for the Coma cluster. The dashed blue lines are the 68\%, 95\% and 99.9\% confidence limits obtained from the WMAP 7-year measurements of the SZ radial profile in the $V+W$ band. The red dotted lines are the confidence limits obtained from the SZ measurements towards the centre of the cluster in different frequency bands first presented in \cite{Davis09}. The black solid lines are the combined confidence limits from these two data sets.}
\end{figure}

The results of this maximum likelihood procedure are presented in Fig. \ref{Fig:CLs}. The blue dashed line shows the 68\%, 95\% and 99\% confidence limits on the parameter space of $\Delta T_{\rm SZ,RJ}(0)$ and $\Delta T_{\rm CSZ}^{204\,\mathrm{GHz}}(0)$. Considering only these WMAP radial measurements we then find that, 
\begin{eqnarray*}
\Delta T_{\rm CSZ}^{204\,\mathrm{GHz}}(0) &=& -60\pm25\,\mu {\rm K} \\
\Delta T_{\rm SZ,RJ}(0) &=& -480\pm100\,\mu {\rm K}\;\;\;(68\% \text{CL}). 
\end{eqnarray*}
In addition to this data set, the SZ measurements towards the centre of the Coma cluster given in Table \ref{tab:ComaSZ} were analysed by the authors in \cite{Davis09}. The resulting constraints on the parameter space were, 
\be
\Delta T_{\rm CSZ}^{204\,\mathrm{GHz}}(0) &\gtrsim & -20\,\mu {\rm K} \nonumber\\
\Delta T_{\rm SZ,RJ}(0) &=& -420\pm60\,\mu {\rm K}\;\;\;(68\% \text{CL}),\nonumber 
\ee
and are reproduced in Fig. \ref{Fig:CLs} as the red dotted line set. Combining these two data sets we find,
\be
\Delta T_{\rm CSZ}^{204\,\mathrm{GHz}}(0) &=& -20\pm15\,\mu {\rm K} \nonumber\\
\Delta T_{\rm SZ,RJ}(0) &=& -400\pm40\,\mu {\rm K}\;\;\;(68\% \text{CL}),\nonumber 
\ee
where the 68\%, 95\% and 99.9\% confidence limits on the parameter space are also plotted in Fig. \ref{Fig:CLs}. We note that, assuming no CSZ contribution, Komatsu \textit{et al.} \cite{Komatsu10} inferred $\Delta T_{\rm SZ,RJ}(0) = -377\pm105\,\mu\mathrm{K}$.  

\begin{figure*}[htb!]
\begin{centering}
\includegraphics[width = 5cm]{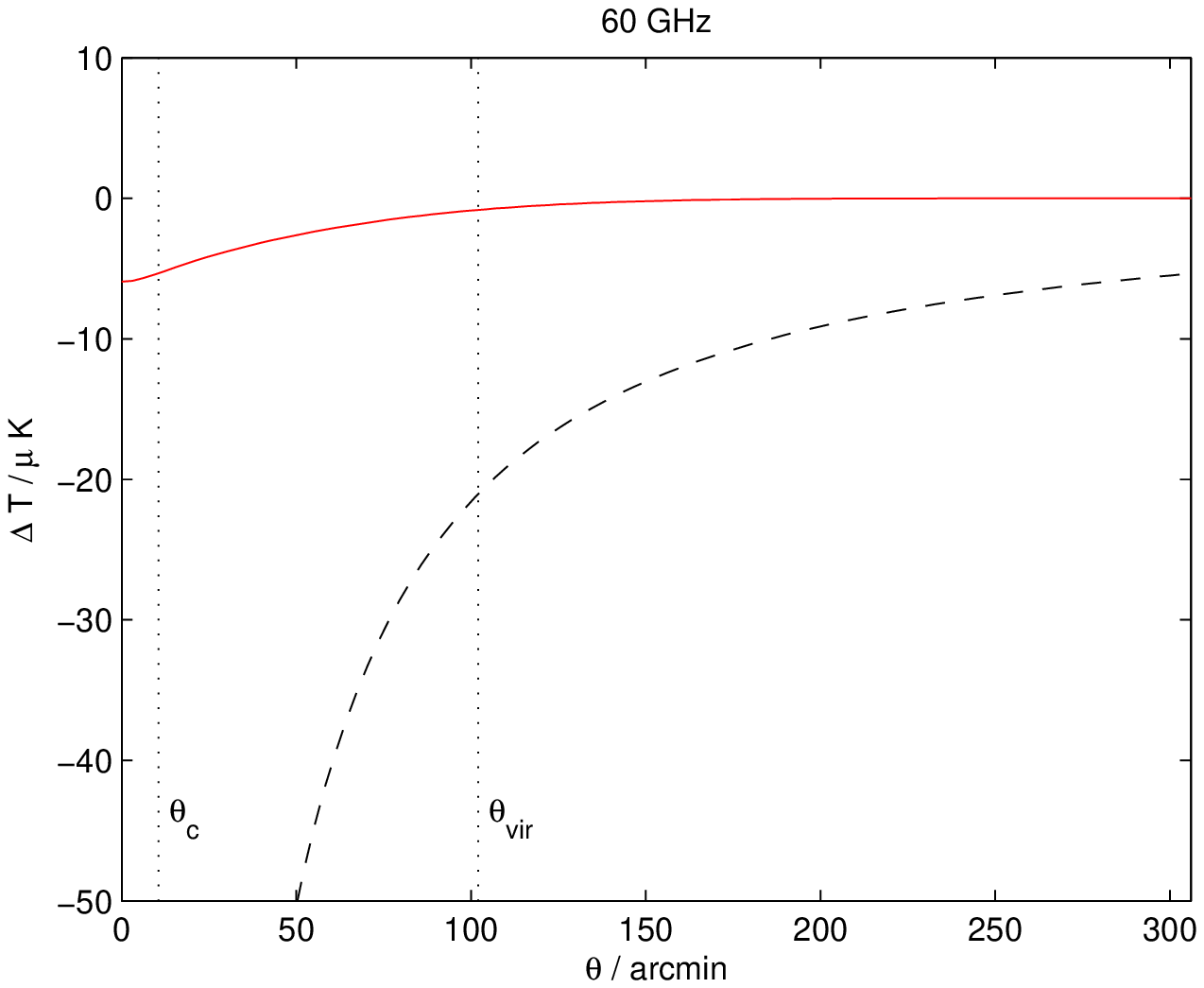}\includegraphics[width = 5cm]{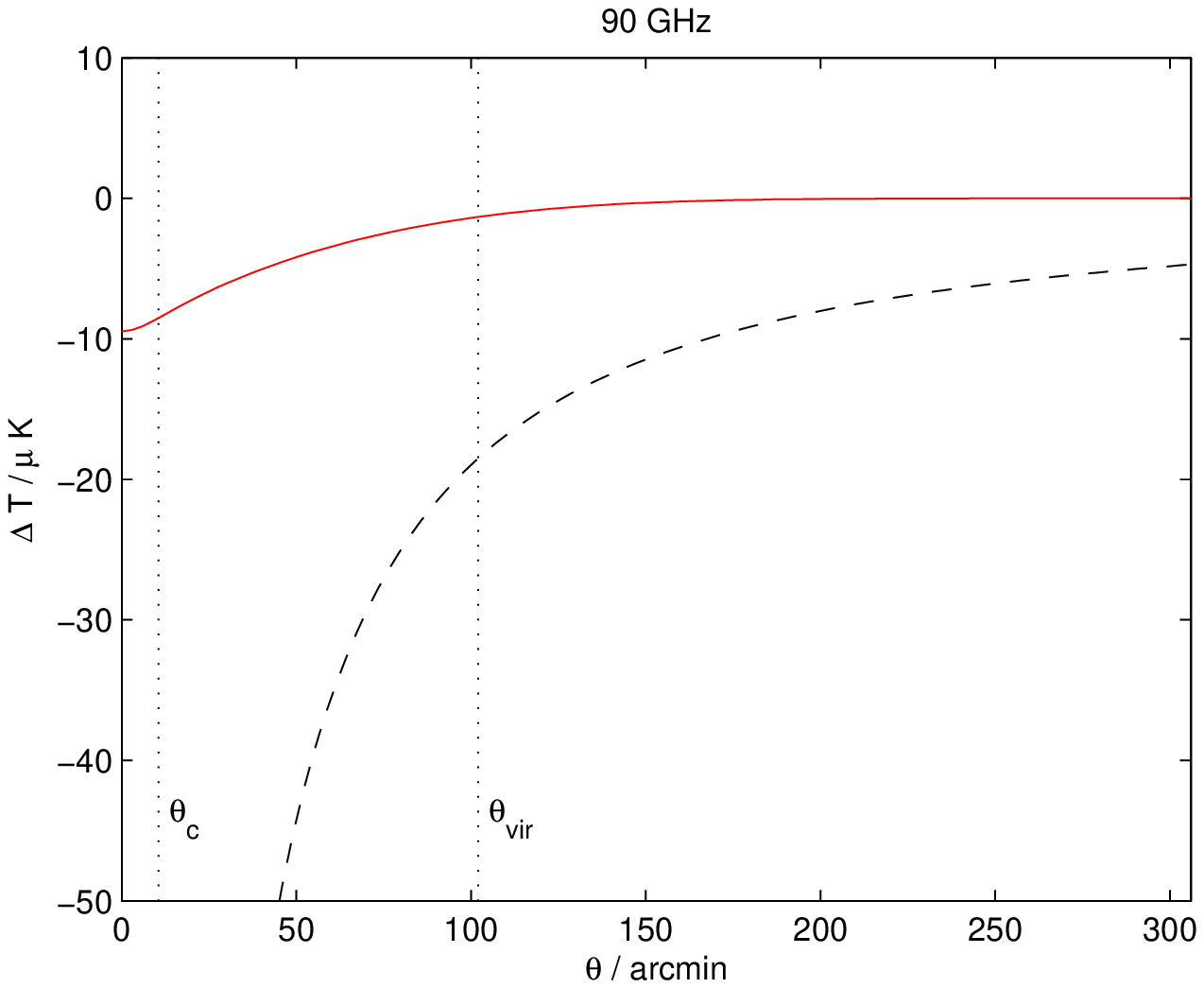}\includegraphics[width = 5cm]{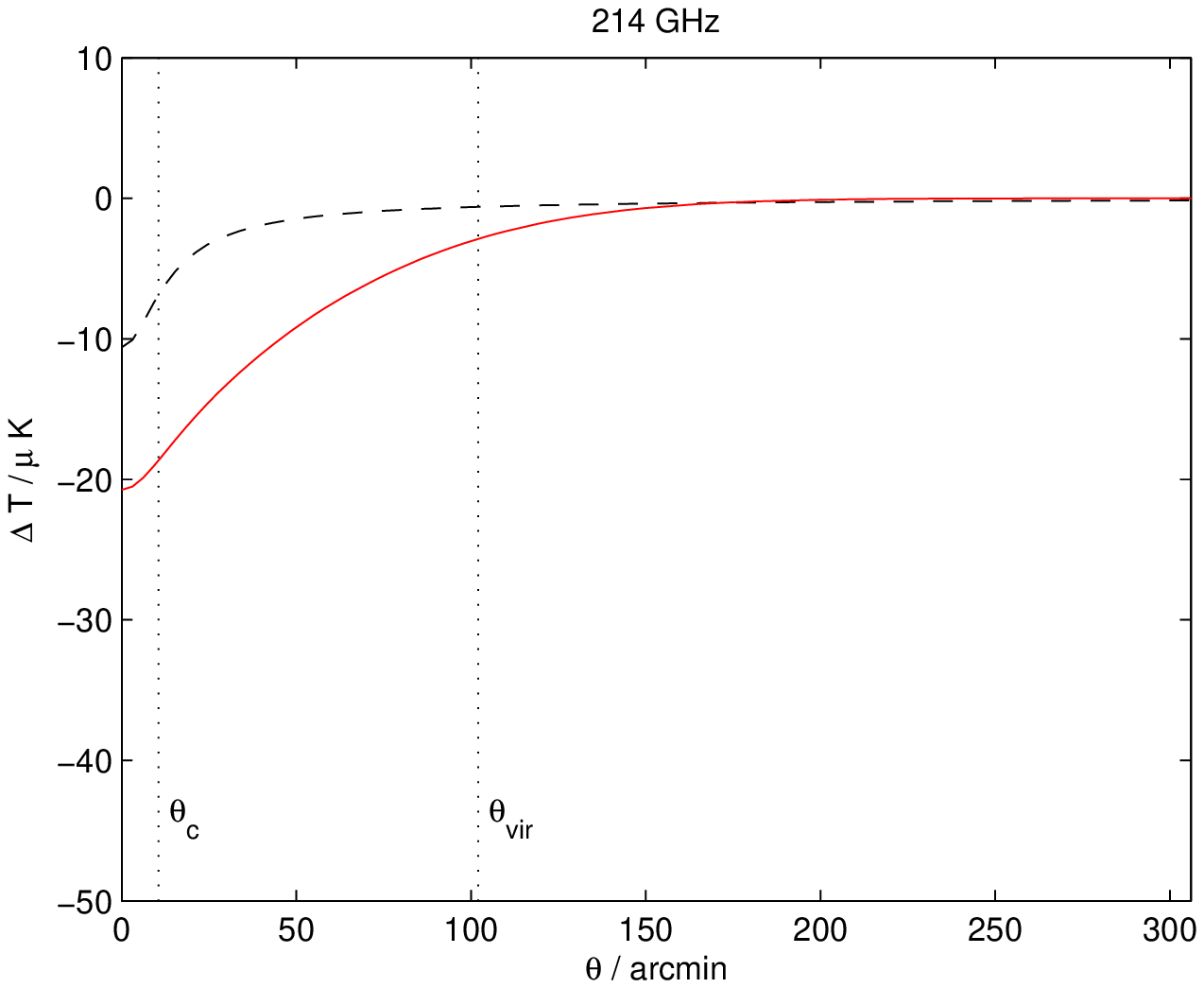}
\par\end{centering}

\caption{\label{fig:profile}The predicted contributions from the CSZ (solid red line) and thermal SZ (dashed black line) effects in the Coma cluster at three different frequencies. We have taken the central temperature decrements to be those of the maximum likelihood parameters, $\Delta T_{\rm SZ,RJ}(0) = -400\mu\mathrm{K}$ and $\Delta T_{\rm CSZ}^{\rm 204GHz}=-20\mu\mathrm{K}$, and have assumed $\theta_{\rm c} \approx 10.5\,\mathrm{arcmin}$, $\gamma = \beta \approx 0.75$, $\eta \approx 0.9$ and $\theta_{\rm vir} \approx 102\,{\rm arcmin}$.}

\end{figure*}

The CSZ contribution to the WMAP $V+W$ band, implied by $\Delta T_{\rm CSZ}^{204\,\mathrm{GHz}}(0)=-20\,\mu {\rm K}$, is plotted in Fig. \ref{Fig:ComaRP}. In Fig. \ref{fig:profile} we have plotted the expected contributions to the signal at $60\,\mathrm{GHz}$, $90\,\mathrm{GHz}$ and $214\,\mathrm{GHz}$. The thermal SZ effect is approximately null at $214\,\mathrm{GHz}$ and so future measurements at this frequency will be most affected by the CSZ contribution. 

The maximum likelihood estimate, $\Delta T_{\rm CSZ}^{204\,\mathrm{GHz}}(0)=-20\,\mu {\rm K}$, corresponds to a photon-scalar conversion probability in the Coma cluster of $\Pbar (204\,\mathrm{GHz}) = 2.7\times 10^{-5}$. Substituting into Eq. (\ref{P204}), we find an implied coupling strength between photons and axion-like particles of, 
\be
g_{\rm eff} \approx (5.2 \text{--} 23.8) \times 10^{-10}\,{\rm GeV}^{-1}. \nonumber
\ee
Similarly the 95\% confidence bound on the CSZ contribution from the combined data set, $\Delta T_{\rm CSZ}^{204\,\mathrm{GHz}}(0) > -55\,\mu {\rm K}$, corresponds to $\Pbar (204\,\mathrm{GHz}) < 7.5\times 10^{-5}$, giving a bound on the photon-scalar coupling strength of,
\be
g_{\rm eff} \lesssim (8.7 \text{--} 39.4) \times 10^{-10}\,{\rm GeV}^{-1}\text{ (95\% CL)}. \nonumber
\ee

\section{Discussion\label{sec:Conclusions}}
In this work we have constrained the degree of conversion between photons and ALPs in the magnetic field of the Coma cluster. The existence of the ALP induces a chameleonic SZ effect which was first predicted in Ref. \cite{Davis09}. Measurements of the SZ effect in the direction of the Coma cluster have been made by a number of CMB probes. Considering only the measured radial profile of the SZ effect in the combined $V+W$ band, we find $\Delta T_{\rm CSZ}^{204\,\mathrm{GHz}}(0)= -60\pm25\,\mu {\rm K}$, for the change to the CMB temperature due to the CSZ effect towards the centre of the cluster. Considering instead SZ measurements towards the centre of the cluster in different frequency bands, the constraint is $\Delta T_{\rm CSZ}^{204\,\mathrm{GHz}}(0)\gtrsim -45\,\mu {\rm K}$ at 95\% confidence. Combining these results we find, 
\be 
\Delta T_{\rm CSZ}^{204\mathrm{GHz}}(0)&\approx&-20\pm15\,\mu\mathrm{K}\text{ (68\% CL)},\nonumber \\
\text{and } \Delta T_{\rm CSZ}^{204\mathrm{GHz}}(0)&\gtrsim&-55\,\mu\mathrm{K}\text{ (95\% CL)}.\nonumber
\ee

The above result involves a number of assumptions. Most importantly, it assumes that the electron density distribution in the cluster can be modelled by a simple beta profile (Eq. \ref{beta_model}), common in both the measurements of cluster surface brightness and SZ profiles. However measurements of the stacked SZ profiles around clusters have suggested that this is a poor fit to the data and that a more sophisticated model is required \cite{ROSATvsWMAP,Komatsu10}. 

As was pointed out in \cite{Davis09}, estimates of the corresponding photon to scalar (or pseudoscalar) coupling strength is highly dependent on the model that is assumed for the cluster magnetic field. In this analysis we have allowed for a spectrum of fluctuations in the magnetic field and electron density and assumed that fluctuations on small scales can be approximated by three-dimensional Kolmogorov turbulence. This leads to an estimate of the photon to scalar coupling strength for the maximum likelihood scenario of, $$g_{\rm eff} \approx (5.2 \text{--} 23.8) \times 10^{-10}\,{\rm GeV}^{-1},$$ while for the 95\% confidence bound on the coupling strength is, $$g_{\rm eff} \lesssim (8.7 \text{--} 39.4) \times 10^{-10}\,{\rm GeV}^{-1}.$$ The range depends on the magnitude of electron density fluctuations in the cluster.

In our analysis we have also imposed that the ALP mass be less than $10^{-12}\mathrm{eV}$. Combining this with the above coupling strength, we find that a standard ALP with those properties is ruled out, since it would violate constraints
on ALP production in the Sun, He burning stars and SN1987A.  However, these constraints feature ALP production in high density
regions, and therefore do not apply to a chameleon scalar field. This includes both the chameleon discussed earlier and any other non-standard ALP for which one or both
of $g_{\rm eff}$ and $1/m_{\phi}$ decrease strongly enough as the
ambient density increases. The above estimate and bound on the chameleon-photon coupling strength is consistent with constraints that have been derived elsewhere \cite{Burrage08,BurrageSN,Davis09}. 

In our radial analysis of the Coma cluster we calculated the CSZ contribution to the $V+W$ band extracted from the WMAP data. To confirm our findings, future work should include an analysis of the raw CMB data in the separate $V$ and $W$ bands. As we can see from Fig. \ref{fig:profile}, the CSZ effect is predicted to dominate over the thermal SZ effect at $214\,\mathrm{GHz}$ and so measurements at these higher frequencies will be most sensitive to the existence of a chameleon. 

We conclude by noting that the anomalous dip in the WMAP measurements at large radii that can be seen in Fig. \ref{Fig:ComaRP} cannot be attributed to photon-scalar mixing: extending the window function that imposed rapid decay of the CSZ contribution beyond the virial radius, merely decreases the total SZ contribution uniformly and provides a poorer fit to the data.

\vspace{0.5cm}

\noindent{\bf Acknowledgments:} ACD, CAOS and DJS are supported by
STFC. We thank Anthony Challinor and Eiichiro Komatsu for helpful comments on a draft version of this manuscript, and are grateful to Eiichiro Komatsu for providing the SZ radial measurements of the Coma cluster in the combined V+W band.

\appendix
\section{ALP-Photon Conversion in Cluster Magnetic Fields\label{appendix:A1}}
In this section we present the calculations for the probability of conversion between photons and ALPs in the magnetic field of galaxy clusters given in Eq. (\ref{Pbar}). The bulk of this derivation was first presented in \cite{Davis09}. We assume a Kolmogorov-type turbulence model for the magnetic field structure in the cluster and allow for similar fluctuations in the electron density. 

Starting from the mixing equations for scalars and pseudoscalars, Eqns. (\ref{eq:mixing1}) and (\ref{eq:mixing2}) given in section \ref{sec:ALPmixing}, we consider a photon field propagating in the $\hatb{z}$ direction, so
that in an orthonormal Cartesian basis $(\hatb{x}, \hatb{y},
\hatb{z})$ we have $\mathbf{a} = (\gamma_x, \gamma_y, 0)^{\rm T}$. The equations of motion for the scalar-ALP and photon polarization
states can then be written in matrix form: 
\be
\left[ -\partial_{t}^{2} +\partial_{z}^{2} - \ba \omega_{\rm pl}^{2} &
  0 & -\frac{B_{y}}{M_{\rm eff}}\partial_{z} \\ 0 & \omega_{\rm pl}^{2}
  & \frac{B_{x}}{M_{\rm eff}}\partial_{z}  \\ \frac{B_{y}}{M_{\rm
      eff}}\partial_{z} & -\frac{B_{x}}{M_{\rm eff}}\partial_{z} &
  m_{\phi}^{2} \ea \right] \bv \gamma_{x} \\ \gamma_{y} \\ \varphi \ev
= 0, \nonumber
\ee 
while for a pseudoscalar-ALP the mixing matrix is
\be
\left[ -\partial_{t}^{2} +\partial_{z}^{2} - \ba \omega_{\rm pl}^{2} &
  0 & -\frac{B_{x}}{M_{\rm eff}}\partial_{t} \\ 
  0 & \omega_{\rm pl}^{2} & -\frac{B_{y}}{M_{\rm eff}}\partial_{t}  \\ 
  \frac{B_{x}}{M_{\rm eff}}\partial_{t} & \frac{B_{y}}{M_{\rm eff}}\partial_{t} &
  m_{\phi}^{2} \ea \right] \bv \gamma_{x} \\ \gamma_{y} \\ \varphi \ev
= 0, \nonumber
\ee 
where we have written $g_{\rm eff}=1/M_{\rm eff}$. 
Following a similar procedure to that in Ref. \cite{Raffelt88}, we
assume the fields vary slowly over time and that the refractive index
is close to unity which requires $m_{\rm \phi}^2/2\omega^2$, $\omega_{\rm
  pl}^2/2\omega^2$ and $\vert B\vert /2\omega M_{\rm eff}$ all
$\ll 1$. Defining $\gamma_{i} = \tilde{\gamma}_{i}(z)
e^{i\omega(z-t)+i \beta(z)}$ and $\varphi = \tilde{\varphi}
e^{i\omega(z-t)+i \beta(z)}$ where $\beta_{,z} = -\omega_{\rm
  pl}^2(z)/2\omega$,  we approximate $-\partial_{t}^{2}\approx
\omega^{2}$ and $\omega^2+\partial_{z}^{2}\approx 2\omega (\omega
+i\partial_{z})$. In addition to this, we define the state vector
\be
\mathbf{u} = \bv \tilde{\gamma}_{x}(z) \\ \tilde{\gamma}_{y}(z) \\ e^{2i\Delta(z)}\tilde{\varphi}(z)\ev. \nonumber
\ee
where $$\Delta(z) =  \int_{0}^{z} \frac{m_{\rm eff}^2(x)}{4\omega} \dd x, $$
and $m_{\rm eff}^2 = m_{\phi}^2(z) - \omega_{\rm pl}^2(z)$, which
simplifies the above mixing field equations for $\mathbf{a}$ and
$\varphi$ to 
\be
\mathbf{u}_{,z} = \frac{\mathcal{B}(z)}{2M_{\rm eff}} \mathbf{u}, \label{stateEqn}
\ee
where
\be
\mathcal{B}(z) = \ba 0 & 0 & -B_{y}e^{-2i\Delta} \\ 0 & 0  &B_{x}e^{-2i\Delta} \\ B_{y}e^{2i\Delta} & -B_{x}e^{2i\Delta} & 0 \ea, \nonumber
\ee
in the case of scalar-photon mixing, and 
\be
\mathcal{B}(z) = \ba 0 & 0 & B_{x}e^{-2i\Delta} \\ 0 & 0  &B_{y}e^{-2i\Delta} \\ -B_{x}e^{2i\Delta} & -B_{y}e^{2i\Delta} & 0 \ea, \nonumber
\ee
in the case of pseudoscalar-photon mixing.

To solve this system of equations we expand, $\mathbf{u} =
\mathbf{u}_{0}+\mathbf{u}_{1}+\mathbf{u}_{2}+\ldots$, such that
\be
{\mathbf{u}}_{i+1}' = \frac{\mathcal{B}(z)}{2M_{\rm
    eff}}\mathbf{u}_{i}\,;\;{\mathbf{u}}_{0}' =0\,, \nonumber
\ee
and neglect the higher order terms from mixing. We define 
$$\mathcal{M}_{1} = \int_{0}^{z} \frac{\mathcal{B}(x)}{2M_{\rm eff}} \dd x. $$
We say that mixing is weak when
\be
{\rm tr}\left( \mathcal{M}_{1}^{\dagger}(z)\mathcal{M}_{1}(z) \right) \ll 1. \nonumber
\ee
Thus Eq. (\ref{stateEqn}) can be solved in the limit of weak-mixing: 
\be
\mathbf{u}(z) \simeq \left[\mathbb{I}+\mathcal{M}_{1}(z) + \mathcal{M}_{2}(z)\right] \mathbf{u}(0), \nonumber
\ee
where $ \mathcal{M}_{2} =
\int_{0}^{z}\mathcal{M}_{1}^{\prime}(x)\mathcal{M}_{1}(x) \dd x$, which
can be written
\be
\mathbf{u}(z) \simeq \left[\mathbb{I}+\mathcal{M}_{1}(z) + \frac{1}{2}\left(\mathcal{M}_{C}(z)+\mathcal{M}_{1}^2(z)\right)\right] \mathbf{u}(0), \nonumber
\ee
where explicitly in the case of scalar-photon mixing
\be
\mathcal{M}_{1} &=& \ba 0 & 0 & -A_{y}^{\ast} \\ 0 & 0 & A_x^{\ast} \\ A_y & -A_x & 0 \ea, \nonumber \\
\mathcal{M}_{C} &=& \ba -C_{yy} & -C^{\ast}_{xy} & 0 \\ C_{xy} & -C_{xx} & 0 \\  0 & 0 & C_{xx}+C_{yy} \ea, \nonumber
\ee
and 
\be
A_{i} &=& \int_{0}^{z} \dd x \frac{B_{i}(x)e^{2i\Delta(x)}}{2M_{\rm eff}}, \label{Ai} \\ 
C_{ij} &=& \int_{0}^{z} \dd x\, \left( A_{i}^{\ast \prime}(x)A_{j}(x) - A_{i}^{\ast}(x)A_{j}^{\prime}(x)\right). \nonumber
\ee
The same holds for pseudoscalar-photon mixing with the replacements $A_x\rightarrow A_y$, $A_y\rightarrow -A_x$, $C_{xx}\leftrightarrow C_{yy}$ and $C_{xy}\rightarrow -C_{yx}$. 

The polarization state of radiation is described by its Stokes
parameters: intensity $I_{\gamma}(z) = \vert\gamma_{x}(z)\vert^2 +
\vert \gamma_{y}(z)\vert^2$, linear polarization $Q(z) =
\vert\gamma_{x}(z)\vert^2 - \vert \gamma_{y}(z)\vert^2$ and $U(z) =
2{\rm Re} \left(\gamma_{x}^{\ast}(z)\gamma_{y}(z)\right)$, and
circular polarization $V(z) = 2{\rm Im}
\left(\gamma_{x}^{\ast}(z)\gamma_{y}(z)\right)$. Assuming there is no initial chameleon flux, $I_{\phi} = \vert \varphi\vert^2 = 0$, then to leading order the final photon intensity is given by
\be
I_{\gamma}(z) &=& I_{\gamma}(0)\left(1-\Pphi(z)\right) \nonumber \\ &&+ Q(0) \mathcal{Q}_{\rm q}(z) + U(0) \mathcal{Q}_{\rm u}(z) + V(0) \mathcal{Q}_{\rm v}(z), \nonumber
\ee
where we have defined,
\be
\Pphi(z) &=& \frac{1}{2}\left(\vert A_{x}(z)\vert^2 + \vert A_{y}(z)\vert^2\right), \nonumber \\
\mathcal{Q}_{\rm q}(z) &=& \frac{1}{2}\left(\vert A_{x}(z)\vert^2 - \vert A_{y}(z)\vert^2\right), \nonumber \\
\mathcal{Q}_{\rm u}(z) &=& {\rm Re}(A_{x}^{\ast}A_{y}), \nonumber \\
\mathcal{Q}_{\rm v}(z) &=& {\rm Im}(A_{x}^{\ast}A_{y}).\nonumber
\ee
For the CSZ effect, we are concerned with CMB radiation, the intrinsic polarization of which is small,
i.e. $\sqrt{Q^2(0)+U^2(0)+V^2(0)}/I_{\gamma}(0) \ll 1$.  It follows
that to leading order the modification to the intensity of
the CMB radiation, from the effect of both scalar- and pseudoscalar-photon mixing, is
\be
\Delta I_{\gamma}(z) &\approx& -\Pphi(z) I_{\gamma}(0). \nonumber
\ee
We now proceed to evaluate the $A_i$ assuming a Kolmogorov-type turbulence model for the magnetic field and electron density fluctuations. 

The magnetic field is split into a regular component, coherent on the scale of the cluster, and a fluctuating component which undergoes field reversals on a much smaller scale, $\mathbf{B}(z)= \mathbf{B}_{\rm reg}+ \delta\mathbf{B}(z)$. We assume that, for each $i$
 and for fixed $\mathbf{x}$, the turbulent component of the magnetic
 field $\delta B_{i}(\mathbf{x})$ is approximately a Gaussian
 random variable. We require (at least approximate) position independence for the
 fluctuations so that $R_{{\rm  B}\,{ij}}(\mathbf{x};\mathbf{y}) \approx
 R_{{\rm B}\,{ij}}(\mathbf{x})$, where $R_{{\rm  B}\,{ij}}(\mathbf{x};\mathbf{y})\equiv\langle\delta B_i (\mathbf{y})\delta B_j (\mathbf{x}+\mathbf{y})\rangle$. $R_{{\rm B}\,ij}(\mathbf{x})$ is the auto-correlation function for
 the magnetic field fluctuations. 
 
We require that $\omega_{\rm pl}^2 \gg m_{\phi}^2$, so that 
$m_{\rm eff}^2 \approx -\omega_{\rm pl}^2 \propto -n_{\rm e}$. Spatial variations in $m_{\rm eff}^2$ are then entirely due to spatial variations in $n_{\rm e}$. The electron number density is divided into a constant part and a fluctuating part, $n_{\rm e} = \bar{n}_{\rm e} + \delta
 n_{\rm e}$, and we assume that $1+\delta n_{\rm e}/\bar{n}_{\rm e}$  is a
 log-normally distributed random variable with mean $1$ and variance
 $\left\langle \delta_{\rm n}^2 \right\rangle$, where $\delta_{\rm
 n}\equiv \delta n_{\rm e}/\bar{n}_{\rm e}$.  We define the electron
 density auto-correlation function by, $$R_{\rm N}(\mathbf{x}) = \left\langle \delta n_{\rm e}(\mathbf{y})\delta n_{\rm e}(\mathbf{x}+\mathbf{y})\right\rangle.$$  
We assume isotropy, which implies $R_{{\rm N}}(\mathbf{x}) = R_{{\rm N}}(x)$. Similarly, isotropy allows us to write the magnetic field auto-correlation function in terms of two scalar correlation functions: 
\be 
R_{{\rm B}\,ij}(\mathbf{x}) &=& \frac{1}{3}R_{\rm B}^{\perp}(x) \delta_{ij} + \hat{x}_{i}\hat{x}_{j} R_{\rm B}^{\parallel}(x), \nonumber
\ee
where $\hatb{x}$ is the unit vector in the direction of $\mathbf{x}$. For example, taking $\mathbf{x}$ in the $z$-direction gives $\hatb{x}=(0,0,1)$. We note that ${\rm Tr}\left(R_{{\rm B} \, ij}(\mathbf{x})\right)=R_{\rm B}^{\perp}(x) + R_{\rm B}^{\parallel}(x)=\left\langle \delta \mathbf{B}(\mathbf{y})\cdot\delta \mathbf{B}(\mathbf{x}+\mathbf{y})\right\rangle$. 
The magnetic field must obey $\nabla \cdot \mathbf{B}=0$ which implies $\partial_{i}R_{{\rm B}\,ij}(\mathbf{x}) =0$ and hence,
\be 
\frac{1}{3}\frac{\dd R_{\rm B}^{\perp}(x)}{\dd x} + \frac{2}{x}  R_{\rm B}^{\parallel}(x) + \frac{\dd R_{\rm B}^{\parallel}(x)}{\dd x} = 0, \nonumber
\ee
and we must have $R_{\rm B}^{\parallel}(0)=0$.  This fixes $R_{\rm B}^{\parallel}(x)$ in terms of $R_{\rm B}^{\perp}(x)$.

The power spectra for the  magnetic and electron density fluctuations, $P_{\rm B}(k)$ and $P_{\rm N}(k)$, are defined:
\be
R_{\rm B}^{\perp}(x) &=& \frac{1}{4\pi} \int \dd^3 k e^{2\pi i \mathbf{k}\cdot\mathbf{x}} P_{\rm B}(k) \nonumber \\
&=& \int k^2 \dd k \,P_{\rm B}(k) {\rm sinc}(2\pi k x), \nonumber \\
R_{\rm N}(x) &=& \frac{1}{4\pi} \int \dd^3 k e^{2\pi i \mathbf{k}\cdot\mathbf{x}} P_{\rm N}(k) \nonumber \\
&=& \int k^2 \dd k \,P_{\rm N}(k) {\rm sinc}(2\pi k x), \nonumber 
\ee
where ${\rm sinc}(x) =\sin x/x$.

We also define,
$$
\bar{\delta}_{\rm n}(z)=\frac{1}{z}\int_{0}^{z} \delta_{\rm n}(x) \dd x\,.
$$
Note $\bar{\delta}_{\rm n}(L)=0$,where $L$ is the path length through the magnetic region. With $A_{i}$ defined by
Eq. (\ref{Ai}), we define $Z = (1+\bar{\delta}_{\rm n}(z))z$ and
$\bar{\Delta} = \bar{m}_{\rm eff}^2L/4\omega\propto -\bar{n}_{\rm e}L /4\omega$. It
follows that $\Delta(z) = \bar{\Delta}Z/L$.  We assume that
$\vert\bar{\delta}_{\rm n}(z)\vert \ll 1$ along the majority of the path, and
so $B_{i}(z) \approx B_{i}(Z)$. We then have,
\be
A_{i}(L) \approx \int_{0}^{L}  \dd Z\,\frac{B_{i}(Z\hatb{z})}{2M_{\rm eff}(1+\delta_{\rm n}(Z\hatb{z}))}e^{\frac{2i\bar{\Delta}Z}{L}}. \nonumber
\ee
We define the correlation,
\be
&\mathcal{A}_{ij} \equiv \left\langle A_{i}(L)A_{j}^{\ast}(L)\right\rangle &\nonumber\\
&= \frac{1}{4M_{\rm eff}^2}\int_{0}^{L}\dd x\int_{0}^{L}\dd y R_{ij}^{\rm tot}((x-y)\hatb{z})e^{2i\bar{\Delta}(x-y)/L}&, \nonumber
\ee
where the correlation function is defined,
$$
R_{ij}^{\rm tot}(x\hatb{z}) \equiv \left\langle \frac{B_{i}(y\hatb{z})B_{j}((y+x)\hatb{z})}{(1+\delta_{\rm n}(y\hatb{z}))(1+\delta_{\rm n}((y+x)\hatb{z}))}\right\rangle.
$$
We split the magnetic field into a regular and turbulent component:
$\mathbf{B}=\mathbf{B}_{\rm reg}+\delta\mathbf{B}$. Thus,
\be
R_{ij}^{\rm tot}(x\hatb{z}) &=& B^{\rm reg}_{i}B^{\rm reg}_{j} \nonumber \\ &&\cdot\left\langle (1+\delta_{\rm n}(y\hatb{z}))^{-1}(1+\delta_{\rm n}((y+x)\hatb{z}))^{-1}\right\rangle \nonumber \\
&&+\left\langle \frac{\delta B_{i}(y\hatb{z}) \cdot \delta B_{j}((y+x)\hatb{z})}{(1+\delta_{\rm n}(y\hatb{z}))(1+\delta_{\rm n}((y+x)\hatb{z}))}\right\rangle. \nonumber
\ee
For simplicity, we assume that fluctuations in the magnetic field and electron density are uncorrelated. Thus,
\be
&\left\langle  \frac{\delta B_{i}(\mathbf{y}) \delta B_{j}(\mathbf{y}+\mathbf{x})}{(1+\delta_{\rm n}(\mathbf{y}))(1+\delta_{\rm n}(\mathbf{x}+\mathbf{y}))}\right\rangle 
= \left\langle \delta B_{i}(\mathbf{y}) \delta B_{j}(\mathbf{y}+\mathbf{x})\right\rangle & \nonumber\\ 
&\cdot\left\langle (1+\delta_{\rm n}(\mathbf{y}))^{-1}(1+\delta_{\rm n}(\mathbf{y}+\mathbf{x}))^{-1}\right\rangle. &\nonumber
\ee
Defining,
\be
R_{\delta}(\mathbf{x})&=&\left\langle (1+\delta_{\rm n}(\mathbf{y}))^{-1}(1+\delta_{\rm n}(\mathbf{y}+\mathbf{x}))^{-1}\right\rangle, \nonumber
\ee
and assuming isotropy we have, 
\be 
R_{ij}^{\rm tot}(x\hatb{z}) =  \left(B^{\rm reg}_{i}B^{\rm reg}_{j} + \frac{1}{3} \delta_{ij} R_{\rm B}^{\perp}(x)  + \hat{z}_{i}\hat{z}_{j} R_{\rm B}^{\parallel}(x)\right) R_{\delta}(x), \nonumber
\ee
where $\hatb{z}=(0,0,1)$.
The photon-scalar conversion probability is given by $\Pphi = \frac{1}{2}\left(\vert A_{x}\vert^2+\vert A_{y}\vert^2\right) = \frac{1}{2}( \hatb{z}\times \mathbf{A})^{\dagger} ( \hatb{z}\times \mathbf{A})$ where $\mathbf{A} = \left(A_{x}(L),A_{y}(L),A_{z}(L)\right)^{\rm T}$. Thus the average conversion probability depends only on the components of $\mathbf{A}$, and hence $R_{ij}^{\rm tot}(x\hatb{z})$, that are perpendicular to $\hatb{z}$. We therefore define,
\be 
\bar{R}_{ij}^{\rm tot}(x\hatb{z}) &=& B^{\rm reg}_{i}B^{\rm reg}_{j} R_{\delta}(x) + \frac{1}{3} \delta_{ij} R_{\rm B}^{\perp}(x)R_{\delta}(x), \nonumber 
\ee
and take, 
\be
\mathcal{A}_{ij} &=& \frac{1}{4M_{\rm eff}^2}\int_{0}^{L}\dd x\int_{0}^{L}\dd y \bar{R}_{ij}^{\rm tot}((x-y)\hatb{z})e^{2i\bar{\Delta}(x-y)/L}. \nonumber
\ee
Since $R_{B}^{\parallel}(x)$ must vanish when $x=0$, we have
\be
\bar{R}_{ij}^{\rm tot}(0) = R_{ij}^{\rm tot}(0) = \left(B^{\rm reg}_{i}B^{\rm reg}_{j}+\frac{\delta_{ij}}{3}\left\langle\vert \delta \mathbf{B}(y\hatb{z})\vert^2\right\rangle\right) R_{\delta}(0). \nonumber
\ee
We define a combined power spectrum $P^{\rm tot}_{ij}(k)$ by,
\be 
\bar{R}_{ij}^{\rm tot}(x\hatb{z})= \int k^2 \dd k\, P^{\rm tot}_{ij}(k) {\rm sinc}(2\pi k x). \nonumber
\ee
We then have,
\be
\mathcal{A}_{ij} = \frac{L^2}{4M_{\rm eff}^2}\int_{0}^{\infty} k^2 P_{ij}^{\rm tot}(k)K(\pi kL;\bar{\Delta})\dd k,\nonumber
\ee
where, 
\be
K(\pi kL;\bar{\Delta}) = \int_{0}^{1} \dd s\; 2(1-s)\frac{\sin(2\pi kL s)}{2\pi kL s}
\cos (2\bar{\Delta}s). \nonumber
\ee
This integral may be evaluated explicitly in terms of the function,
\be
\Si(x) &=& \int_{0}^{x} \frac{\sin y}{y} \dd y.\nonumber
\ee
We define $\delta \Si(x,y) = \Si(x+y)-\Si(x-y)$.   We then have,
\be
2xK(x;\bar{\Delta}) &=& \delta\Si(2\bar{\Delta},2x) \nonumber\\
&& + \frac{\bar{\Delta}\sin 2\bar{\Delta}\sin 2x + x\cos 2\bar{\Delta}\cos 2x-x}{x^2 - \bar{\Delta}^2}. \nonumber 
\ee
We are concerned with CMB radiation, for which $\omega \sim
10^{-5}\text{--}10^{-3}\eV$, propagating over distances of the order
of $100\kpc$ through a galaxy cluster.  It is easy to see that for such a scenario $\vert\bar{\Delta}\vert \gg 1$.  We evaluate $K(x;\bar{\Delta})$ in the asymptotic
 limit where $\vert \bar{\Delta}\vert \gg 1$. When $x \ll \vert \bar{\Delta}\vert$ we have,
\be
4xK(x;\bar{\Delta}) &\simeq & \frac{2x}{\bar{\Delta}^2} - \frac{\sin 2x \cos 2\bar{\Delta}}{\bar{\Delta}^2},\nonumber
\ee
and when $x \gg \vert \bar{\Delta}\vert$, we have
\be
4xK(x;\bar{\Delta}) &\simeq& 2\pi - \frac{2}{x}.\nonumber
\ee 

We define $k_{\rm crit} = \vert \bar{\Delta} \vert/\pi L$. We assume that
the fluctuations in $\mathbf{B}$ and $n_{\rm e}$ are such that $P^{\rm tot}_{ij}(k)$ drops off faster than $k^{-3}$ for all $k >k_{\ast}$ where $k_{\ast} \ll k_{\rm crit}$. 
This is equivalent to assuming that the dominant contribution to $R^{\rm tot}_{ij}(0)$ comes from spatial scales than are much larger than $k_{\rm crit}^{-1}$.  We expect that this 
dominant contribution will come from scales of the order of the coherence lengths of the magnetic field and electron density fluctuations ($L_{\rm B}$ and $L_{\rm N}$ defined in Eq. (\ref{lengths})) and so we are assuming that $k_{\rm crit}^{-1} \ll L_{\rm B},\,L_{\rm N}$.   With this assumption, in the limit of large $\vert \bar{\Delta}\vert$, we have to leading 
order,
\be
4M_{\rm eff}^2\mathcal{A}_{ij} &\simeq& \frac{L^2}{2\bar{\Delta}^2} \int_{0}^{\infty} k^2 P_{ij}^{\rm tot}(k)\dd k \nonumber \\
&& - \frac{L \cos 2\bar{\Delta}}{4\pi \bar{\Delta}^2} \int_{0}^{\infty} k \sin(2\pi kL) P_{ij}^{\rm tot}(k)\dd k \nonumber \\
&& + \frac{L}{2} \int_{k_{\rm crit}}^{\infty} k P_{ij}^{\rm tot}(k)\dd k \,.\nonumber
\ee
We define,
$$
W_{ij}(k) = \int_{k}^{\infty} qP_{ij}^{\rm tot}(q) \dd q, 
$$
and we can then rewrite $\mathcal{A}_{ij}$ as,
\be
\mathcal{A}_{ij} &\simeq& \frac{2 \omega^2}{M_{\rm eff}^2 \bar{m}_{\rm eff}^4}\bar{R}_{ij}^{\rm tot}(0) - \frac{2 \omega^2}{M_{\rm eff}^2 \bar{m}_{\rm eff}^4} \bar{R}_{ij}^{\rm tot}(L\hatb{z}) \cos 2\bar{\Delta} \nonumber \\
&&+\frac{L}{8M_{\rm eff}^2} W_{ij}(k_{\rm crit}). \nonumber
\ee
We have that,
\be
\bar{R}_{ij}^{\rm tot}(0) = R_{ij}^{\rm tot}(0) \approx \left(B_{i}^{\rm reg}B_{j}^{\rm reg} + \frac{1}{3}\delta_{ij}\left\langle \delta \mathbf{B}^2\right\rangle\right)\left\langle
\frac{\bar{n}_{\rm e}^2}{n_{\rm e}^2}\right\rangle. \nonumber
\ee
We assume that fluctuations in $n_{\rm e}$ follow an approximately log-normal distribution, and hence
\be
\left\langle\frac{\bar{n}_{\rm e}^2}{n_{\rm e}^2}\right\rangle \approx\left\langle\frac{n_{\rm e}^2}{\bar{n}_{\rm e}^2}\right\rangle^3 \equiv I_{\rm N}^3\,. \nonumber
\ee
We assume, as is the case in the situations we consider, that $L$ is much larger than the correlation length of the magnetic and electron density fluctuations.  
This implies that the only contribution to $\bar{R}^{\rm tot}_{ij}(L\hatb{z})$ which may result in a leading order contribution to $\mathcal{A}_{ij}$ is,
\be
\bar{R}_{ij}^{\rm tot}(L\hatb{z}) \sim B_{i}^{\rm reg}B_{j}^{\rm reg}. \nonumber
\ee
Finally, we must evaluate $W_{ij}(k)$ in terms of the magnetic and electron density power spectra. 
We separate the fluctuations in $n_{\rm e}$  into short and long wavelength fluctuations, $\delta_s$ and $\delta_l$ respectively, and 
assume they are approximately independent. Thus $n_{\rm e}=\bar{n}_{\rm e}(1+\delta_l)(1+\delta_s)$.  This parameterisation is consistent with the assumption that $n_{\rm e}/\bar{n}_{\rm e}$ has an 
 approximately log-normal distribution. We assume that the short wavelength fluctuations are linear up to some cut-off scale 
 $k_{\rm lin}^{-1}$. Above this spatial scale we have the long wavelength fluctuations, which are not necessarily linear. We then have that over small spatial scales, $\ll k^{-1}_{\rm lin}$,
\be
R_{\delta}(\mathbf{x}) &\approx& \left\langle \frac{\bar{n}_{\rm e}^2}{n_{\rm e}^2}\right\rangle \left\langle (1+\delta_{s}(\mathbf{y}))^{-1}(1+\delta_{s}(\mathbf{y}+\mathbf{x}))^{-1}\right\rangle,\nonumber  \\
&\approx& \left\langle \frac{\bar{n}_{\rm e}^2}{n_{\rm e}^2}\right\rangle\left\langle \frac{n_{\rm e}^2}{\bar{n}_{\rm e}^2}\right\rangle^{-1} \left[1+ n_{\rm e}^{-2} R_{\rm N}(\mathbf{x})\right], \nonumber \\
&=& I_{\rm N}^2 \left[ 1 + \bar{n}_{\rm e}^{-2}R_{\rm N}(\mathbf{x})\right]. \nonumber
\ee
It follows that for $k \gg k_{\rm lin}$ we have approximately,
\be
P_{\delta}(k) \approx I_{\rm N}^2 \bar{n}_{\rm e}^{-2} P_{\rm N}(k). \nonumber
\ee
We assume that $k_{\rm lin} \ll k_{\rm crit}$.  Finally we define,
\be
P_{\rm B\delta}(k) = P_{\rm B\delta}(\mathbf{k}) &=& \frac{1}{4\pi} \int \dd^3 p P_{\delta}(p)P_{\rm B}(\Vert \mathbf{k}-\mathbf{p}\Vert),\nonumber 
\ee
which is the Fourier transform of $R_{\rm B}^{\perp}(x)R_{\delta}(x)$.
From the definition of $P_{ij}^{\rm tot}$ and $W_{ij}(k)$ we have,
\be
W_{ij}(k) &=& B_{i}^{\rm reg} B_{j}^{\rm reg}\int_{k}^{\infty}q P_{\delta}(q)\dd q \nonumber\\
&&+\frac{1}{3}\delta_{ij}\int_{k}^{\infty}q P_{B\delta}(q)\dd q. \nonumber
\ee
All that remains is to evaluate,
\be
\int_{k}^{\infty} q P_{\rm B\delta}(q) \dd q = \int \dd^3 p \frac{P_{\delta}(p)}{4\pi} \int_{V_{k}(\mathbf{p})} \dd^3 r \frac{P_{\rm B}(r)}{4\pi \Vert \mathbf{r}+\mathbf{p}\Vert}, \nonumber
\ee
where $V_{k}(\mathbf{p})$ is defined by $\Vert \mathbf{r}+\mathbf{p}\Vert > k$.  We approximate this integral using $\Vert \mathbf{r} + \mathbf{p}\Vert 
\approx p$ in $0 < r < p$ and $\Vert \mathbf{r} + \mathbf{p}\Vert \approx r$ in $0 < p < r$, and find
\be
\int_{k}^{\infty} q P_{B\delta}(q)\dd q &\approx& \left\langle \delta \mathbf{B}^2 \right\rangle \int_{k}^{\infty}q P_{\delta}(q)\dd q \nonumber \\ &&+ I_{N}^3 \int_{k}^{\infty}q
 P_{B}(q)\dd q. \nonumber
\ee
Defining,
\be
W_{\rm B/N}(k) = \int_{k}^{\infty} q\dd q P_{\rm B/N}(q), \nonumber
\ee
we have for $\mathcal{A}_{ij}$,
\be
\mathcal{A}_{ij} &\simeq& \frac{2 I_{N}^3 B_{ij}^2 \omega^2}{M_{\rm eff}^2 \bar{m}_{\rm eff}^4} - \frac{2 B_{i}^{\rm reg} B_{j}^{\rm reg} \omega^2}{M_{\rm eff}^2 \bar{m}_{\rm eff}^4}\cos 2\bar{\Delta} \nonumber \\
&&+\frac{I_{\rm N}^2 L}{8M_{\rm eff}^2}\left(\frac{B_{ij}^2}{\bar{n}_e^2} W_{N}(k_{\rm crit})+ \frac{\delta_{ij} I_{\rm N}}{3} W_{B}(k_{\rm crit})\right), \nonumber
\ee
where $B^{2}_{ij} = B_{i}^{\rm reg}B_{j}^{\rm reg} + \delta_{ij}\left\langle \delta \mathbf{B}^2\right\rangle /3$. It follows that,
\be
\Pbar &\equiv& \frac{1}{2}\left(\mathcal{A}_{xx}+\mathcal{A}_{yy}\right) \nonumber\\
&\approx & \frac{1}{2}\left(\frac{2B_{\rm eff} \omega}{M_{\rm eff} \bar{m}_{\rm eff}^2}\right)^2 I_{\rm N}^3 - \frac{1}{4}\left(\frac{2B_{\rm reg} \omega}{M_{\rm eff} \bar{m}_{\rm eff}^2}\right)^2 \cos \left(2\bar{\Delta}\right) \nonumber 
\\ && + \frac{B_{\rm eff}^2 L }{8M_{\rm eff}^2 \bar{n}_{\rm e}^2} I_{\rm N}^2 W_{\rm N}(k_{\rm crit}) + \frac{L}{24M_{\rm eff}^2} I_{\rm N}^3  W_{\rm B}(k_{\rm crit}), \nonumber
\ee
where $B_{\rm eff}^2 \equiv B_{\rm reg}^2/2 + \left\langle \delta \mathbf{B}^2\right\rangle/3$. 
The term that scales as $L$ is associated with electron density fluctuations and is usually dominant when $\vert \bar{\Delta}\vert \gg 1$, as is the case for CMB photons in galaxy clusters.

We have assumed that near $k=k_{\rm crit}$ and for all large $k$, both $P_{\rm B}(k)$ and  $P_{\rm N}(k)$ are decreasing faster than $k^{-3}$. We have
\be
k_{\rm crit}^{-1} \approx 2.4 \times 10^{-2}\pc \left(\frac{\nu}{100\,{\rm GHz}}\right) \left(\frac{10^{-3}\,{\rm cm}^{-3}}{\bar{n}_{\rm e}}\right), \nonumber
\ee
where $\nu = \omega/2\pi$ is the frequency of the electromagnetic
radiation.  For galaxy clusters  $\bar{n}_{\rm e} \sim 10^{-3}\textrm{--} 10^{-2}\,{\rm cm}^{-3}$ and for CMB photons $\nu \sim
30-300\,{\rm GHz}$. The typical critical length scale is then much smaller than a parsec, $k_{\rm crit}^{-1} 
\approx 10^{-3}\textrm{--}0.1\,{\rm pc}$. The form and magnitude of $P_{\rm B}(k)$ and $P_{\rm N}(k)$ in galaxy clusters for such small scales are not known empirically. We assume that near 
$k = k_{\rm crit}$, $P_{\rm B}(k) \propto P_{\rm N}(k) \propto k^{-11/3}$ corresponding to three-dimensional Kolmogorov turbulence. We define normalization constants $C_{\rm K}$ and $C_{\rm N}$,
\be
k^2 P_{\rm B}(k) &=& 2C_{\rm K}\left(\frac{k}{k_{0}}\right)^{-5/3}, \nonumber\\
k^2 P_{\rm N}(k) &=& 2(2\pi)^{1/3} C_{\rm N}^2 k^{-5/3}, \nonumber
\ee
for consistency with Refs. \cite{Han04,Davis09}, where $k_{0} = 1\kpc^{-1}$. It follows that,
\be
W_{\rm B}(k_{\rm crit}) &\approx & \frac{6C_{\rm K}}{5} \left(\frac{k_{\rm crit}}{k_{0}}\right)^{-5/3}, \nonumber \\
W_{\rm N}(k_{\rm crit}) &\approx & \frac{6(2\pi)^{1/3}}{5}  C_{\rm N}^2 k_{\rm crit}^{-5/3}. \nonumber
\ee
Order of magnitude estimates for the normalisation constants were calculated in \cite{Davis09}:
\be
2k_{0}^{5/3}C_{\rm K}  &\approx& \left(0.27 \textrm{--} 0.45\right)\left\langle \delta \mathbf{B}^2 \right\rangle L_{\rm B}^{-2/3} \nonumber \\
2(2\pi)^{1/3}C_{\rm N}^2 \bar{n}_{e}^{-2} &\approx& \left(0.27 \textrm{--} 0.45\right)\frac{\left\langle\delta n_{\rm e}^2\right\rangle}{\bar{n}_{\rm e}^2} L_{\rm B}^{-2/3}, \nonumber
\ee
where $\left\langle\delta n_{\rm e}^2\right\rangle/\bar{n}_{\rm e}^2 =I_{\rm N}-1$, and $L_{\rm N}$ is approximated by $L_{\rm B}$.

Thus we find that the dominant contribution to the photon-ALP conversion rate is, 
\be
\Pbar &\approx& \frac{3}{40}(0.27\text{--}0.45) \left(\frac{\alpha_{\rm EM}\bar{n}_{\rm e}}{2\pi m_{\rm e}}\right)^{-5/3} \nu^{5/3}L_{\rm B}^{-2/3}M_{\rm eff}^{-2} L \nonumber \\
&& \cdot\left(B_{\rm reg}^{2}\frac{I_{\rm N}^2\left(I_{\rm N}-1\right)}{2}+ \left\langle \delta \mathbf{B}^2\right\rangle \frac{I_{\rm N}^2 \left(2I_{\rm N}-1\right)}{3}\right). \nonumber \\
&& \label{eq:Pbar}
\ee

\section{CSZ Contribution to the WMAP Signal Extracted from the $V+W$ Bands\label{appendix:A2}}

In this appendix, we explain our simplistic approach to the processing technique of Komatsu \textit{et al.} in combining the raw data from the WMAP $V$ and $W$ bands in order to remove contamination from the CMB anisotropies \cite{Komatsu10}.

Considering only the contribution from
the thermal SZ effect and CMB anisotropies, the observed temperature decrement around the cluster is expected to be,
\[
\Delta T(\theta,\nu)= \Delta T_{CMB}(\theta)+ \Delta T_{SZ}(\theta,\nu), \]
where the fluctuations due to CMB anisotropies are frequency independent.
The frequency dependence of the thermal SZ effect was given in Eq. (\ref{eq:SZ effect}). Subtracting the signal in the $W$ band
from that in the $V$ band we would expect (allowing only for the thermal SZ effect) that, 
\[
\Delta T(\theta,\nu_{V})-\Delta T(\theta,\nu_{W})=\Delta T_{SZ,RJ}(\theta)\left(\frac{g_{\rm SZ}(\nu_{V})-g_{\rm SZ}(\nu_{W})}{-2}\right).\]
The results presented in \cite{Komatsu10} are expected to be of $\Delta T_{\rm RJ}^{V+W}=\Delta T_{\rm SZ,RJ}(\theta)$. 

However if we include the contribution from the CSZ effect, then we expect the total temperature fluctuation in any frequency band to be given by, 
\[
\Delta T(\theta,\nu)=\Delta T_{CMB}(\theta)+\Delta T_{SZ}(\theta,\nu)+\Delta T_{CSZ}(\theta,\nu),\]
where $\Delta T_{\rm CSZ}$ was given in Eq. (\ref{eq:DeltaCSZ}). This leads to a predicted signal of, 
\be
\Delta T_{\rm RJ}^{V+W}(\theta)&=&\Delta T_{\rm SZ,RJ}(\theta) \nonumber \\
&& +\Delta T_{\rm CSZ}^{204}(\theta)\left(\frac{-2(g_{\rm CSZ}^{V}-g_{\rm CSZ}^{W})}{g_{\rm CSZ}(\nu_0)(g_{\rm SZ}^{V}-g_{\rm SZ}^{W})}\right),\nonumber\\
&&\label{combinedT}
\ee
where $g_{\rm CSZ}(\nu)\equiv \nu^{5/3}(e^{-x}-1)/x$ and $\nu_0=204\,\mathrm{GHz}$.

\end{document}